\documentclass[twoside,twocolumn,9pt]{article}
\usepackage{extsizes}
\usepackage[super,sort&compress,comma]{natbib} 
\usepackage[version=3]{mhchem}
\usepackage[left=1.5cm, right=1.5cm, top=1.785cm, bottom=2.0cm]{geometry}
\usepackage{balance}
\usepackage{widetext}
\usepackage{times,mathptmx}
\usepackage{sectsty}
\usepackage{graphicx} 
\usepackage{lastpage}
\usepackage[format=plain,justification=raggedright,singlelinecheck=false,font={stretch=1.125,small,sf},labelfont=bf,labelsep=space]{caption}
\usepackage{float}
\usepackage{fancyhdr}
\usepackage{fnpos}
\usepackage[english]{babel}
\usepackage{array}
\usepackage{droidsans}
\usepackage{charter}
\usepackage[T1]{fontenc}
\usepackage[usenames,dvipsnames]{xcolor}
\usepackage{setspace}
\usepackage{amsmath}
\usepackage[utf8]{inputenc}
\usepackage[compact]{titlesec}
\usepackage{tabu}

\usepackage{epstopdf}

\definecolor{cream}{RGB}{222,217,201}

\makeatletter
\newsavebox{\@brx}
\newcommand{\llangle}[1][]{\savebox{\@brx}{\(\m@th{#1\langle}\)}%
  \mathopen{\copy\@brx\kern-0.5\wd\@brx\usebox{\@brx}}}
\newcommand{\rrangle}[1][]{\savebox{\@brx}{\(\m@th{#1\rangle}\)}%
  \mathclose{\copy\@brx\kern-0.5\wd\@brx\usebox{\@brx}}}
\makeatother

\begin{document}

\pagestyle{fancy}
\thispagestyle{plain}
\fancypagestyle{plain}{

\fancyhead[C]{\includegraphics[width=18.5cm]{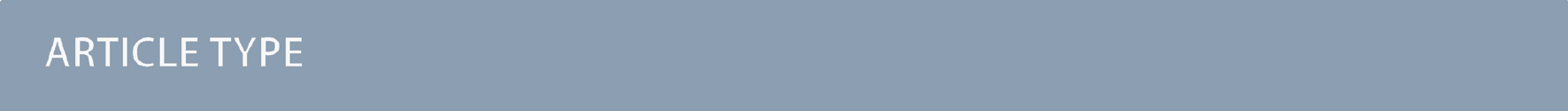}}
\fancyhead[L]{\hspace{0cm}\vspace{1.5cm}\includegraphics[height=30pt]{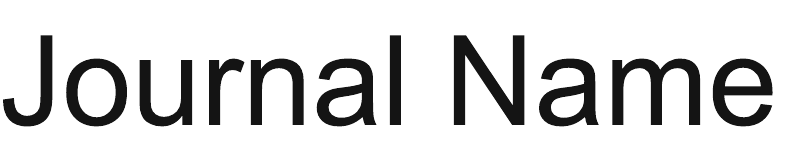}}
\fancyhead[R]{\hspace{0cm}\vspace{1.7cm}\includegraphics[height=55pt]{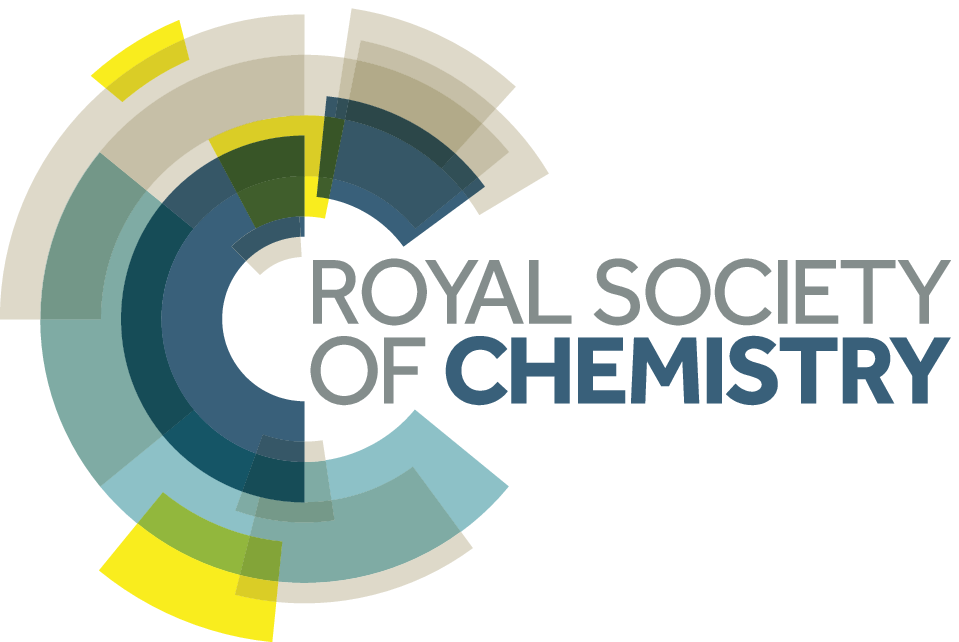}}
\renewcommand{\headrulewidth}{0pt}
}

\makeFNbottom
\makeatletter
\renewcommand\LARGE{\@setfontsize\LARGE{15pt}{17}}
\renewcommand\Large{\@setfontsize\Large{12pt}{14}}
\renewcommand\large{\@setfontsize\large{10pt}{12}}
\renewcommand\footnotesize{\@setfontsize\footnotesize{7pt}{10}}
\makeatother

\renewcommand{\thefootnote}{\fnsymbol{footnote}}
\renewcommand\footnoterule{\vspace*{1pt}%
\color{cream}\hrule width 3.5in height 0.4pt \color{black}\vspace*{5pt}} 
\setcounter{secnumdepth}{5}

\makeatletter 
\renewcommand\@biblabel[1]{#1}            
\renewcommand\@makefntext[1]%
{\noindent\makebox[0pt][r]{\@thefnmark\,}#1}
\makeatother 
\renewcommand{\figurename}{\small{Fig.}~}
\sectionfont{\sffamily\Large}
\subsectionfont{\normalsize}
\subsubsectionfont{\bf}
\setstretch{1.125} 
\setlength{\skip\footins}{0.8cm}
\setlength{\footnotesep}{0.25cm}
\setlength{\jot}{10pt}
\titlespacing*{\section}{0pt}{4pt}{4pt}
\titlespacing*{\subsection}{0pt}{15pt}{1pt}

\fancyfoot{}
\fancyfoot[LO,RE]{\vspace{-7.1pt}\includegraphics[height=9pt]{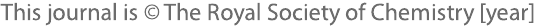}}
\fancyfoot[CO]{\vspace{-7.1pt}\hspace{13.2cm}\includegraphics{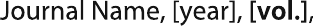}}
\fancyfoot[CE]{\vspace{-7.2pt}\hspace{-14.2cm}\includegraphics{head_foot/RF}}
\fancyfoot[RO]{\footnotesize{\sffamily{1--\pageref{LastPage} ~\textbar  \hspace{2pt}\thepage}}}
\fancyfoot[LE]{\footnotesize{\sffamily{\thepage~\textbar\hspace{3.45cm} 1--\pageref{LastPage}}}}
\fancyhead{}
\renewcommand{\headrulewidth}{0pt} 
\renewcommand{\footrulewidth}{0pt}
\setlength{\arrayrulewidth}{1pt}
\setlength{\columnsep}{6.5mm}
\setlength\bibsep{1pt}

\makeatletter 
\newlength{\figrulesep} 
\setlength{\figrulesep}{0.5\textfloatsep} 

\newcommand{\topfigrule}{\vspace*{-1pt}%
\noindent{\color{cream}\rule[-\figrulesep]{\columnwidth}{1.5pt}} }

\newcommand{\botfigrule}{\vspace*{-2pt}%
\noindent{\color{cream}\rule[\figrulesep]{\columnwidth}{1.5pt}} }

\newcommand{\dblfigrule}{\vspace*{-1pt}%
\noindent{\color{cream}\rule[-\figrulesep]{\textwidth}{1.5pt}} }

\newcommand{\thickhline}{%
    \noalign {\ifnum 0=`}\fi \hrule height 0.5pt
    \futurelet \reserved@a \@xhline
}

\makeatother

\twocolumn[
  \begin{@twocolumnfalse}
\vspace{3cm}
\sffamily
\begin{tabular}{m{4.5cm} p{13.5cm} }

\includegraphics{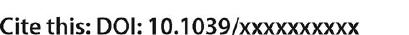} & \noindent\LARGE{\textbf{How to distinguish between interacting and noninteracting molecules in tunnel junctions $^\dag$}} \\
\vspace{0.3cm} & \vspace{0.3cm} \\

& \noindent\large{Miguel A. Sierra,\textit{$^{a}$} David S\'anchez,\textit{$^{a}$} Alvar R. Garrigues\textit{$^{b}$}, Enrique del Barco\textit{$^{b}$}, Lejia Wang\textit{$^{c,d}$} and Christian A. Nijhuis\textit{$^{d,e}$} } \\

\includegraphics{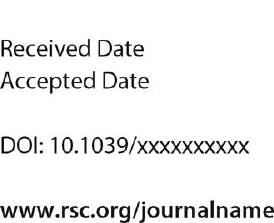} & \noindent\normalsize{%
Recent experiments demonstrate a temperature control of the electric conduction through a ferrocene-based molecular junction. Here we examine the results in view of determining means to distinguish between transport through single-particle molecular levels or via transport channels split by Coulomb repulsion. Both transport mechanisms are similar in molecular junctions given the similarities between molecular intralevel energies and the charging energy. We propose an experimentally testable way to identify the main transport process. By applying a magnetic field to the molecule, we observe that an interacting theory predicts a shift of the conductance resonances of the molecule whereas
in the noninteracting case each resonance is split into two peaks. The interaction model works well in explaining our experimental results obtained in a ferrocene-based single-molecule junction, where the charge degeneracy peaks shift (but do not split) under the action of an applied 7-Tesla magnetic field. This method is useful for a proper characterization of the transport properties of molecular tunnel junctions.} \\

\end{tabular}

 \end{@twocolumnfalse} \vspace{0.6cm}

  ]

\renewcommand*\rmdefault{bch}\normalfont\upshape
\rmfamily
\section*{}
\vspace{-1cm}


\footnotetext{\textit{$^{a}$~Institute for Cross-Disciplinary Physics and Complex Systems IFISC (UIB-CSIC), Palma de Mallorca, Spain. E-mail: david.sanchez@uib.es}}
\footnotetext{\textit{$^{b}$~Department of Physics, University of Central Florida, Orlando, Florida, USA. E-mails: delbarco@ucf.edu}}
\footnotetext{\textit{$^{c}$~School of Chemical Engineering, Ningbo University of Technology, Ningbo, Zhejiang, 315016, P.R. China.}}
\footnotetext{\textit{$^{d}$~Department of Chemistry, National University of Singapore, Singapore. E-mail: christian.nijhuis@nus.edu.sg}}
\footnotetext{\textit{$^{e}$~Centre for Advanced 2D Materials, National University of Singapore, Singapore}}

\footnotetext{\dag~Electronic Supplementary Information (ESI) available: [details of any supplementary information available should be included here]. See DOI: 10.1039/b000000x/}




\section{Introduction}
A molecular tunnel junction comprises molecules trapped between bulk metallic electrodes.
These systems include electromigrated single-electron transistors~\cite{par00,jpa722,yu04,alv115}, scanning tunneling microscopy break-junctions~\cite{ven06,die09}, and self-assembled monolayers of molecules~\cite{sam1,sam2,sam3,sam4,nij611,sam5,sam6,sam7,sam8,sam9}.
The energy spectrum of the molecule is characterized by both the highest occupied molecular
orbital (HOMO) and the lower unoccupied molecular orbital (LUMO). When the HOMO or the LUMO
aligns with the Fermi level of the leads $E_F$ within a certain tunnel broadening $\Gamma$,
charge carriers can resonantly tunnel through the molecule~\cite{Cuevas}. Tuning of the
relative position of the molecular levels with respect to $E_F$ can be effectively
achieved with a capacitively coupled gate electrode.

Since the level spacing between molecular levels $\Delta$
is typically large in molecular junctions, size quantization effects can be observed even at room
temperature provided that $k_B T < \Delta$. However, the transport properties of the tunnel
junction depend on $T$ because the electric current is a function of the electronic occupation at the leads,
which is governed by a Fermi-Dirac distribution. The system response to both temperature and a bias voltage $V$ applied
across the leads has been investigated in different classes of molecules:
molecular diodes~\cite{nij611,yua506,yua324,yoo155}, molecular wires \cite{hoc482,yan326,smi732,laf193,koc713}
and biomolecules\cite{liv040,liw816,sep421,kum824}.

Due to the small size of molecules, in many occasions it is crucial to take into account an additional
energy scale. In incoherent tunneling processes for molecules weakly coupled to the attached metals
(i.e., for junction conductances smaller than $e^2/h$), 
charge carriers spend much time inside
the molecule and Coulomb repulsion effects thus become relevant.
This leads to regions with suppressed transport within the current--voltage characteristics
of the molecular transistor (Coulomb blockade effect)~\cite{thi08}.
Therefore, electrons flow through the molecule only when their energy is larger than
the charging energy $U$. A key signature of Coulomb blockade is the appearance at $k_BT < U$
of areas of forbidden transport
(commonly referred as ``Coulomb diamonds")
in the differential conductance curves as a function of source-drain and gate voltages.
Coulomb blockade has been reported in molecular single-electron transistors~\cite{par00}
as well as in different solid-state systems~\cite{qdots,lee16,leh16,sho17}.

Garrigues \textit{et al.}~\cite{alv115} have recently investigated the temperature dependence of charge transport
across a ferrocene-based molecular transistor. They found that, when $T$ increases, the magnitude of the charge degeneracy points associated to current maxima decreases while the inverse behavior is observed for the current valleys.
A noninteracting theoretical model that takes into account two single-particle levels in the molecule
showed good agreement with the experimental data. Nevertheless, $\Delta$ and $U$ can be of the same order.
Therefore, it is of fundamental importance to determine whether transport is due to noninteracting (independent electrons)
or interacting (Coulomb blockade) physics. In this work, we demonstrate that an interacting theoretical model
can equally well fit the experimental results and hence offer an alternative explanation for the results found
in Ref.~\cite{alv115}.

\begin{figure}[t]
\centering
  \includegraphics[width=0.49\textwidth,clip]{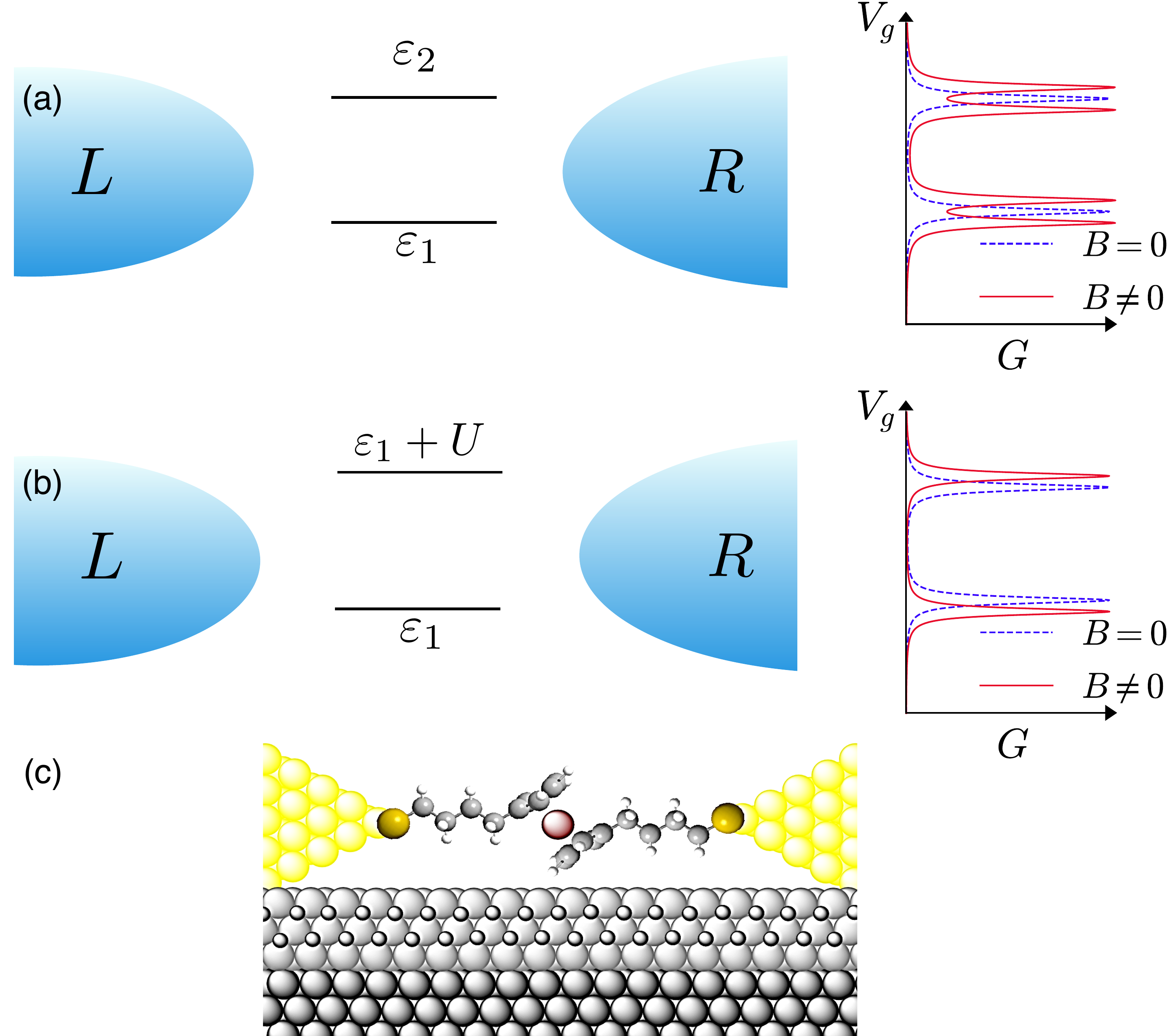}
  \caption{Sketch of our proposal on how to evaluate the importance
  of electron-electron interactions in transport through
  molecules sandwiched between left ($L$) and right ($R$) metallic electrodes (tunnel junction). (a) The noninteracting case considers two molecular levels $\varepsilon_1$ and $\varepsilon_2$, which yield two conductance resonances as shown in the right plot (dashed blue lines). (b) The interacting case encompasses a single level $\varepsilon_1$ and a charging energy $U$, which also give rise to two conductance peaks. Remarkably,
when a magnetic field $B$ is turned on the noninteracting resonances
split as $B$ increases (red curves) whilst the interacting peaks shift.
This different magnetic response allows us to characterize the transport
mechanism in molecular tunnel junctions.
(c) Schematic of the S-(CH$_2$)$_4$-ferrocenyl-(CH$_2$)$_4$-S molecular tunnel junction used to test our proposal.}
  \label{fig:sketch}
\end{figure}

Our finding promptly triggers the following interesting question: How could one distinguish between both transport mechanisms
(noninteracting case and Coulomb blockade) in a particular molecular junction setup? We stress that this question
is unique to molecular devices since quantum dot systems exhibit charging energies much larger than
the mean level spacing~\cite{qdots}.
In molecular tunnel junctions,
the level spacing is comparable to the charging energy,
making it difficult to differentiate between the two cases.
Here, we suggest that an externally applied magnetic field would serve as a tool to discern between
the two transport mechanisms.
Our proposal is illustrated in Fig.~\ref{fig:sketch}. In the top panel
we depict a molecule with two single-particle quantum levels
($\varepsilon_1$ and $\varepsilon_2$) coupled to two electrodes
($L$ and $R$). The linear conductance would then show two peaks
that are spin-split under the action of a magnetic field (top-right panel).
The interacting counterpart is sketched in the lower panel where
we consider one quantum level $\varepsilon_1$ and charging energy $U$.
Due to Coulomb repulsion induced splitting, two conductance peaks
are expected at zero magnetic field (bottom-right panel). Yet,
with increasing magnetic field each resonance is not split but
shifts in energy as shown. The differing current response
would allow us to characterize the precise transport mechanism
in a given molecular tunnel junction. Below, we
support this proposal with a nonequilibrium Green function based calculation that
agrees with the experimentally observed conductance peaks of a ferrocene-based molecular tunnel junction
in the presence of magnetic fields.

\section{Theoretical framework}
Consider a single molecule coupled to
the left, $L$, and right, $R$, metallic electrodes
via tunneling barriers. We model the system with the single-impurity Anderson Hamiltonian,
\begin{equation}\label{eq_anderson}
\mathcal{H}=\mathcal{H}_{\rm mol}+\mathcal{H}_{\rm leads}+\mathcal{H}_{\rm tunnel}\,.
\end{equation}

First, $\mathcal{H}_{\rm mol}$ describes the electrons in the molecule. For definiteness, 
we focus on the $N$-th molecular resonance. Its total energy $E$
is given by the sum of a kinetic part ($\varepsilon_N$) and a potential term.
The latter originates from the interaction with the surrounding electrodes (source, drain and gate $g$
terminals). Following standard electrostatics, we model the coupling between the molecule and the electrodes
with capacitances $C_L$, $C_R$ and $C_g$ (constant interaction model).
Let $\mu_m(N)=E(N)-E(N-1)$ be the molecular electrochemical potential. Then, 
\begin{equation}
\label{Eq:mud}\mu_m(N) = \frac{(2N-1)e^2}{2C}-e \frac{C_LV_L+C_RV_L+C_gV_g}{C}+\varepsilon_N\,,
\end{equation}
($C=C_L+C_R+C_g$ is the total capacitance). Calculation of $\mu_m(N)$ is important because it determines
the molecule's addition energy, $\Delta\mu(N) = \mu_m(N+1) - \mu_m(N)$. It is straightforward to see that
$\Delta\mu_m(N) =\Delta_N + e^2 /C$, where $\Delta_N = \varepsilon_{N+1} - \varepsilon_{N}$.
For the moment, we consider the spin-degenerate case (we will later discuss magnetic field effects).
Then, $\varepsilon_{N+1} = \varepsilon_{N}$ and the resonance spacing is entirely given
by the charging energy $U=e^2/C$. As a consequence, we can write
\begin{align}\label{eq_hmol}
\mathcal{H}_{\rm mol}=\sum_\sigma \varepsilon_m d^\dagger_\sigma d_\sigma
+ U d_{\uparrow}^\dagger d_\uparrow d_\downarrow^\dagger d_\downarrow \,,
\end{align}
where $\varepsilon_m = \varepsilon_N -e (C_LV_L+C_RV_L+C_gV_g)/C$ is the single-particle energy
including possible level renormalizations due to polarization effects. In Eq.~\eqref{eq_hmol}
$d_\sigma^\dagger$ ($d_\sigma$) is the creation (annihilation) operator for electrons in the localized level
and $\sigma=\{\uparrow,\downarrow\}$.

The metallic electrodes are represented in Eq.~\eqref{eq_anderson} by
\begin{align}
\mathcal{H}_{\rm leads} = \sum_{\alpha k \sigma } \varepsilon_{\alpha k} c_{\alpha k\sigma}^\dagger c_{\alpha k \sigma}\,,
\end{align}
where $\alpha=L,R$ and $c_{\alpha k \sigma}^\dagger$ ($c_{\alpha k \sigma}$) is the creation (annihilation) of conduction electrons with energy spectrum $\varepsilon_{\alpha k}$ ($k$ is the wave number).

Finally, the coupling of the molecule to the electrodes in Eq.~\eqref{eq_anderson} is described by the tunnel Hamiltonian
\begin{align}
\mathcal{H}_{\rm tunnel}=\sum_{\alpha k \sigma} t_{\alpha k}c^\dagger_{\alpha k \sigma} d_\sigma +{\rm h. c.}\, ,
\end{align}
where $t_{\alpha k}$ is the electronic tunnel amplitude.

\section{Electric current}
The current can be found from the time evolution of the expected value of the occupation in one of the electrodes: $I_\alpha = -ed\langle n_\alpha (t)\rangle/dt$, where $n_\alpha = c_{\alpha k \sigma}^\dagger c_{\alpha k \sigma}$. Due to charge conservation in the steady state, one has $I_L+I_R=0$ and this implies that current can be calculated in one of the electrodes only: $I\equiv I_L$. Applying the Keldysh-Green's function formalism \cite{Haug},
the electric current in terms of the transmission function $\mathcal{T}(\omega)$ becomes
\begin{align}\label{Eq:Current}
I=\frac{e}{h} \int_{-\infty}^\infty d\omega \mathcal{T}(\omega) [f_L(\omega)-f_R(\omega)]\,,
\end{align}
with $f_\alpha(\omega) = 1/[\exp{(\omega-\mu_\alpha)/k_BT}+1]$ the leads' Fermi-Dirac distribution with electrochemical potentials $\mu_L = \varepsilon_F +eV/2$ and $\mu_R=\varepsilon_F-eV/2$. The transmission obeys the expression
\begin{align}\label{Eq:Trans}
\mathcal{T}(\omega) = \sum_\sigma {\Gamma}_{L}  {\Gamma}_{R } G^r_{\sigma,\sigma} G^a_{\sigma,\sigma} \, ,
\end{align}
where $\Gamma_{\alpha}(\omega)=2\pi\rho_\alpha |t_{\alpha k}|^2$ is the tunnel hybridization due to coupling with electrode $\alpha$
($\rho_\alpha$ is the density of states for electrode $\alpha$). The total broadening is then $\Gamma =\Gamma_{L}+\Gamma_{R}$.
Quite generally, $\Gamma$ is a function of energy $\omega$ because electron tunneling depends on the position of each molecular level via the tunnel barrier height.
This is important for the fit of the experimental curves, as we will discuss below.

In Eq.~\eqref{Eq:Trans}, $G^r_{\sigma,\sigma}(\omega)$ and $G^a_{\sigma,\sigma}(\omega)$ are the Fourier transforms for the retarded and advanced Green's function, respectively, of the molecular system,
\begin{align}
G^r_{\sigma,\sigma}(t,t')= -\frac{i}{\hbar}\theta(t-t')\langle [d_\sigma(t),d_\sigma^\dagger(t')]_+\rangle \, ,
\end{align}
where $\theta(t-t')$ is the Heaviside function and $[\cdots]_+$ denotes the anticommutator.
\section{Green's function}
In this work we use an
approximate expression that works fairly well for the Coulomb blockade
regime. 
This regime is characterized by strong electron-electron interactions and
weak couplings to the electrodes ($U > k_B T > \Gamma_\alpha$).
Temperature is moderate such that Kondo correlations can be safely
neglected. This amounts to
disregarding both the charge excitation correlators $\llangle d_\sigma d_{\bar{\sigma}}^\dagger C_{\alpha k \bar{\sigma}}, d_\sigma^\dagger\rrangle$, $\llangle d_\sigma C_{\alpha k \bar{\sigma}}^\dagger d_{\bar{\sigma}}, d_\sigma^\dagger\rrangle$ and $\llangle C_{\alpha k\sigma} d_{\bar{\sigma}}^\dagger C_{\alpha k \bar{\sigma}}, d_\sigma^\dagger\rrangle$, and the spin correlator $\llangle C_{\alpha k\sigma} C_{\alpha k \bar{\sigma}}^\dagger d_{\bar{\sigma}}, d_\sigma^\dagger\rrangle $.
Then, the equation-of-motion technique~\cite{mei91} leads to
the following system of equations in the Fourier space
\begin{align}
(\omega-\varepsilon_{m})G^r_{\sigma,\sigma}&=1 +\sum_{\alpha k} t_{\alpha k}^* G^r_{\alpha k\sigma,\sigma} +U \llangle d_\sigma n_{\bar{\sigma}}, d_\sigma^\dagger\rrangle\, , \\
(\omega-\varepsilon_m-U)\llangle d_\sigma n_{\bar{\sigma}}, d_\sigma^\dagger\rrangle  &= \langle n_{\bar{\sigma}} \rangle+\sum_{\alpha k} t_{\alpha k}^* \llangle C_{\alpha k\sigma} n_{\bar{\sigma}}, d_\sigma^\dagger\rrangle \, ,\\
(\omega-\varepsilon_{\alpha k})\llangle C_{\alpha k\sigma} n_{\bar{\sigma}}, d_\sigma^\dagger\rrangle &= t_{\alpha k } \llangle d_{\sigma} n_{\bar{\sigma}}, d_\sigma^\dagger\rrangle \,,
\end{align} 
which can be readily solved for the retarded Green's function
\begin{align}\label{Eq:Gr}
G^r_{\sigma, \sigma}(\omega) = \frac{1-\langle n_{\bar{\sigma}}\rangle }{\omega - \varepsilon_{m} -\Sigma} +\frac{\langle n_{\bar{\sigma}}\rangle }{\omega - \varepsilon_{m} -U -\Sigma} \,,
\end{align}
where $\Sigma\simeq -i\Gamma/2$ and the mean occupation $\langle n_{\bar{\sigma}} \rangle$ of the molecular level
is evaluated from
\begin{align}\label{Eq:nsigma}
\langle n_\sigma \rangle = \int d\omega \frac{\Gamma_{L} f_L(\omega)+\Gamma_{R}f_R(\omega)}{\Gamma} \rho_\sigma(\omega)\,.
\end{align}
$\rho_\sigma(\omega) = -(1/\pi) \text{Im}{[G^r_{\sigma,\sigma}]}$ is the molecular level spectral function.

Interestingly, the Green's function in Eq.~\eqref{Eq:Gr} shows two poles despite the fact that we are considering a single molecular level. These two poles will be resolved in a transport experiment when $U$ is larger than $\Gamma$,
which is the typical situation for weakly coupled molecules. Equation~\eqref{Eq:Gr} is also capable of
describing a strongly coupled molecule (i.e., when $\Gamma\gg U$), in which case the molecule effectively
acts as a noninteracting channel. Therefore, our equation-of-motion model encompasses a broad variety
of situations provided Kondo physics is not important.

Therefore, in the Coulomb blockade regime ($U\gg\Gamma$)
we expect two resonances in the transmission function separated by the charging energy $U$.
This situation resembles very much the case of two molecular levels $\varepsilon_1$ and $\varepsilon_2$
(e.g., HOMO and LUMO) used in Ref.~\cite{alv115}. In fact, the transmission function for two noninteracting levels
is given by the sum of two Breit-Wigner (Lorentzian)
line shapes ($i=1,2$),
\begin{align}\label{Eq:Tlandauer}
\mathcal{T}(\omega)=\sum_i \frac{\gamma_{Li}\gamma_{Ri}}{(\omega-\varepsilon_i)^2+\gamma_i^2/4} \, ,
\end{align}
which also gives rise to two resonances like Eq.~\eqref{Eq:Gr}.
For that reason, it is difficult to tell whether interactions will play
a significant role in the electronic transport through molecules with level spacings
$|\varepsilon_2 - \varepsilon_1|$ of the same order as $U$.
We will below demonstrate that an external magnetic field helps solve this critical issue.

\section{Results}
The self-consistent calculation of Eqs.~\eqref{Eq:Gr} and~\eqref{Eq:nsigma} completely determines the transmission given by Eq.~\eqref{Eq:Trans}.
We will now show that this solution nicely fits the experimental results of Ref.~\cite{alv115}.
\begin{figure}[t]
\centering
  \includegraphics[scale=0.41]{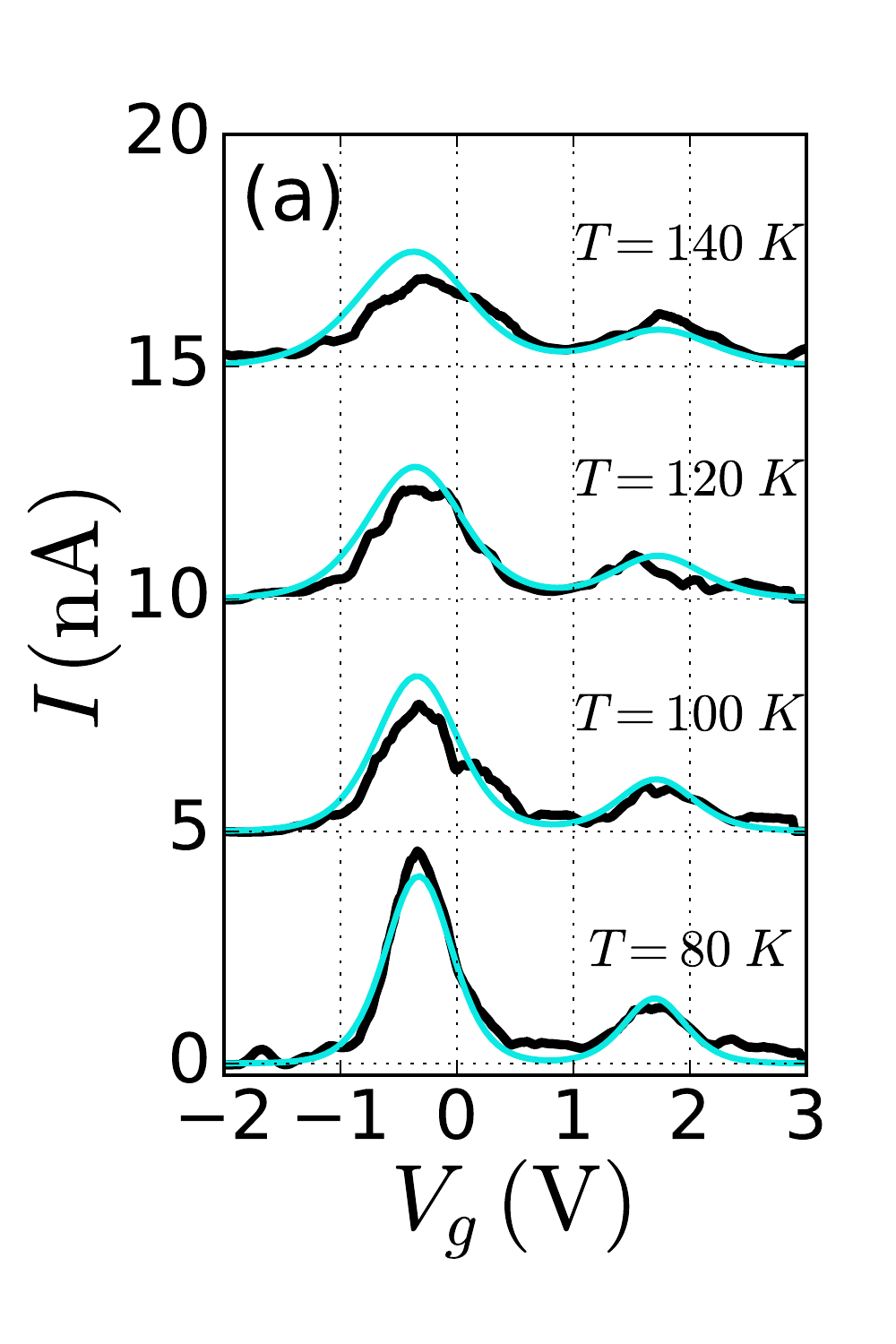}\includegraphics[scale=0.49]{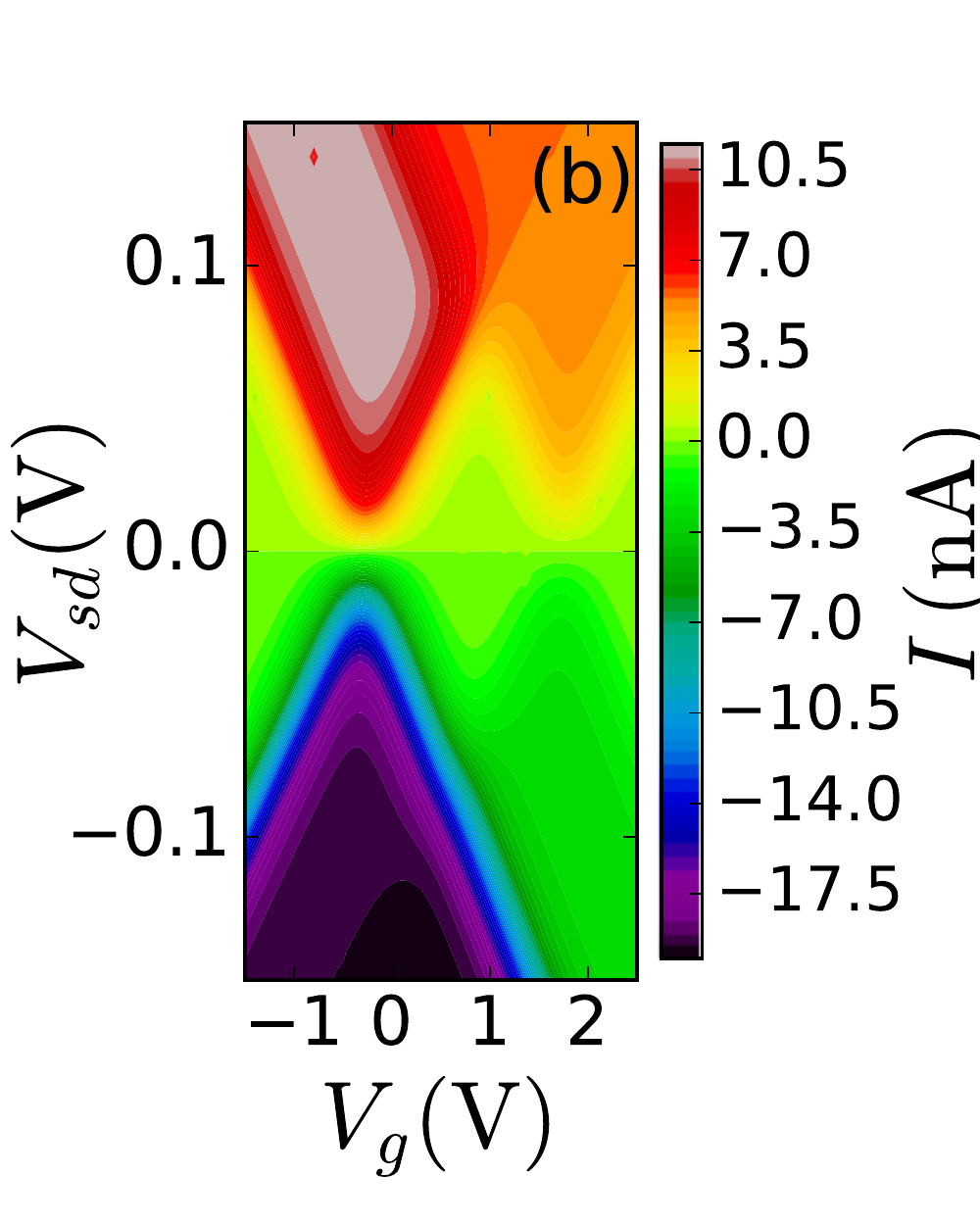}
  \caption{(a) Current $I$ as a function of the gate voltage $V_g$ for different values of the indicated temperatures
  and a source-drain voltage $V=10$~mV. Experimental (theoretical) results are depicted with black (blue) curves
  (experimental data is extracted from Ref.~\cite{alv115}).
  Curves for different temperature are shifted vertically for the sake of clarity (offset: $5$ nA).
  (b) Calculated current at $T= 80$ K as a function of both source-drain
  bias and gate voltage, showing clear Coulomb diamond regions. }
  \label{fig:2}
\end{figure}

We take the following functional dependence of $\Gamma_\alpha$ versus energy:
\begin{align}
\Gamma_\alpha = \begin{cases} \gamma_{\alpha 1} & \text{if} \; \omega < \varepsilon_m+ U/2 \\ \gamma_{\alpha 2} & \text{if} \; \omega > \varepsilon_m + U/2 \end{cases} \, ,
\end{align} 
In Fig.~\ref{fig:2}(a) we plot the experimental data for the measured current
as a function of the gate voltage at different values
of the background temperature (black curves)
for a single S-(CH$_2$)$_4$-ferrocenyl-(CH$_2$)$_4$-S molecule tunnel junction [see the schematic Fig.~\ref{fig:sketch}(c)], as reported in Ref.~\cite{alv115}.
We also show (blue lines) the results
of our theoretical model applying the parameters of Table~\ref{Tab:1}.
(Similar values have been reported elsewhere~\cite{poo06,lea08,xie15}.)
We observe two different peaks in $I(V_g)$. The first peak arises when the molecule energy level aligns the electrochemical potential of one of the leads, $\varepsilon_m\simeq \mu_\alpha$. In contrast to the noninteracting model~\cite{alv115}, the second peak here is not associated with a second molecular level but with a split resonance due to electron-electron interaction [the $U$ resonance in Eq.~\eqref{Eq:Gr}].
With increasing $T$ the height of the peaks smoothly decreases in both the experiment results and the numerical modeling.
This further reinforces the possibility that Coulomb repulsion should be relevant in molecular transport at temperatures $k_BT < U $.
Figure~\ref{fig:2}(b) shows the current as a function of both gate and source-drain voltages for $T=80$~K.
We find clearly visible Coulomb diamonds within which transport is blockaded until $|eV|$ is higher than $U$.
In the limit $|eV| \gg U$ current saturates to the maximum value dictated by the double-barrier resonant-tunneling device.
\begin{table}[t]
  \begin{tabular*}{0.48\textwidth}{@{\extracolsep{\fill}}llll}
    \hline
    $U$  & $\varepsilon_N$ & $\gamma_{L1}$ \\
    \thickhline
    76 meV & 27 meV & 0.4 meV \\
    \hline
    $\gamma_{L2}$ & $\gamma_{R1}$ & $\gamma_{R2}$\\
    \thickhline
    0.4 meV & 0.05 meV & 0.01 meV\\
    \hline
    $C_g$ & $C_L$ & $C_R$ \\
    \thickhline
     0.525 e/V & 5.78 e/V & 6.83 e/V\\
    \hline
  \end{tabular*}
  \caption{Parameters for the molecular transistor used in the theoretical model.}
  \label{Tab:1}
\end{table}

\begin{figure}[t]
\centering
  \includegraphics[scale=0.5]{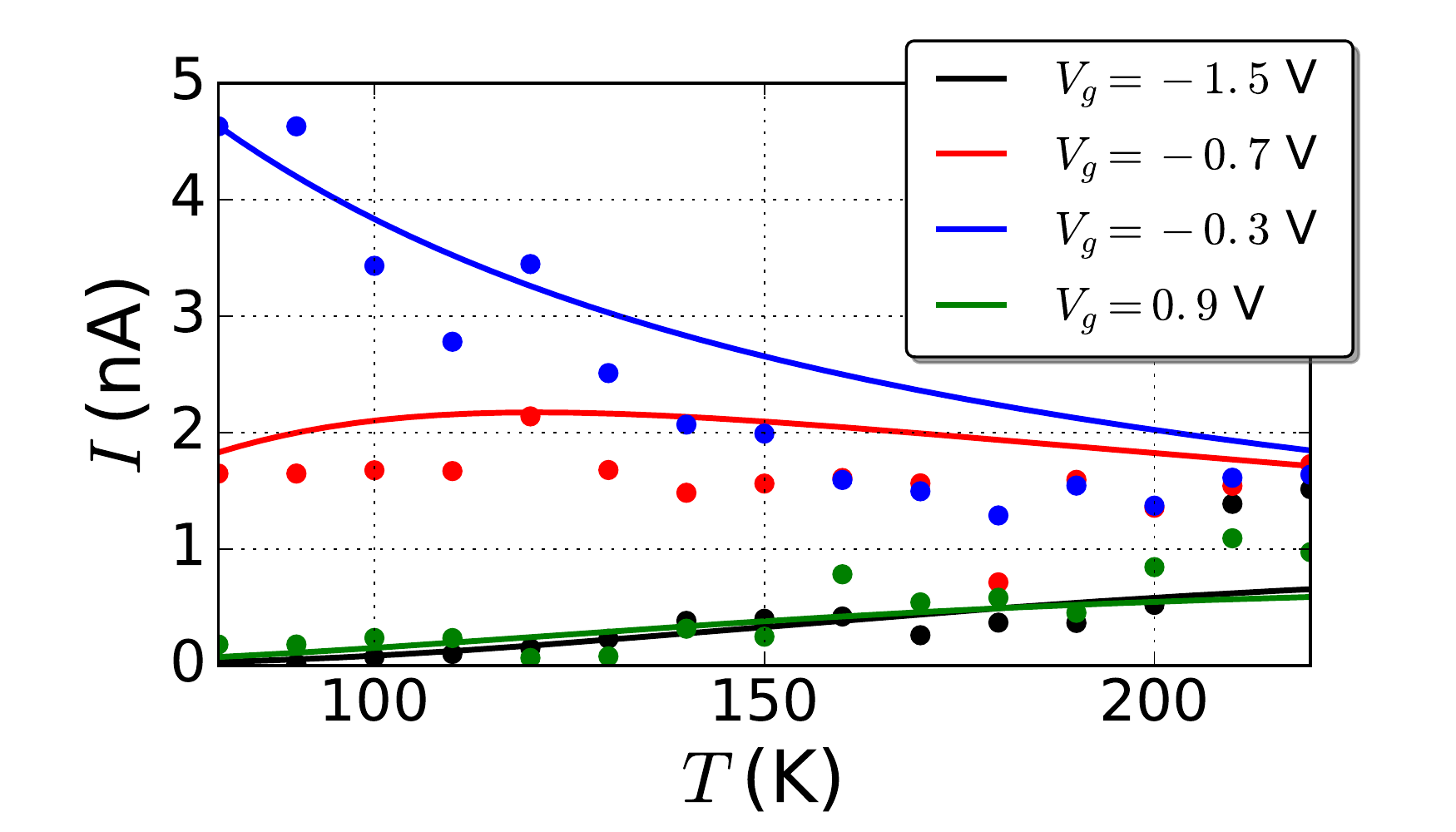}
  \caption{(a) Current at $V_{sd}=10$~meV as a function of the background temperature $T$ for different gate voltage values. Solid lines correspond to our theoretical model while data points show the experimental results of Ref.~\cite{alv115}. We use the model parameters used from Table \ref{Tab:1}}
  \label{fig:3}
\end{figure}

Temperature effects on the current across the molecule are shown in Fig.~\ref{fig:3} for several values of the gate voltage. For $V_g = -0.3$~V, which corresponds to the low-energy peak in Fig.~\ref{fig:2}(a), the current decreases with $T$ as expected from the degeneracy point. When the molecular resonance aligns with the electrode electrochemical potential ($V_g = -0.7$~V), the current is rather insensitive to $T$. Finally,
in the Coulomb blockade valley [to the left ($V_g = -1.5$~V) or the right ($V_g = 0.9$~V) of the current peak] the current in fact increases as $T$ raises. We note that the agreement between theory and experiment is good in all cases (within the experimental error bars, not shown here) except for high temperatures since in this case dephasing and inelastic scattering is more likely to occur in the molecular bridge and our transport theory breaks down.
On the other hand,
our results are valid as long as temperature is not exceedingly small
since higher order co-tunneling processes (e.g., Kondo correlations)
could then take place when the Fc units are strongly hybridized with the electrodes
via, e.g., shortened alkyl side-arms.

\section{Discussion}

We thus have two competing models that describe well the experimental results. In the interacting picture as discussed above, the splitting of the molecular resonance is attributed to Coulomb blockade. In contrast, the noninteracting model employed in Ref.~\cite{alv115} points to two independent molecular resonances. Strikingly, both theories show a temperature-dependent tunnel current compatible with the experiments.
The problem we now face is how to distinguish between these two explanations.
We propose that a magnetic field applied to the junction would constitute a reliable test.

\begin{figure}[t]
\centering
  \includegraphics[scale=0.31]{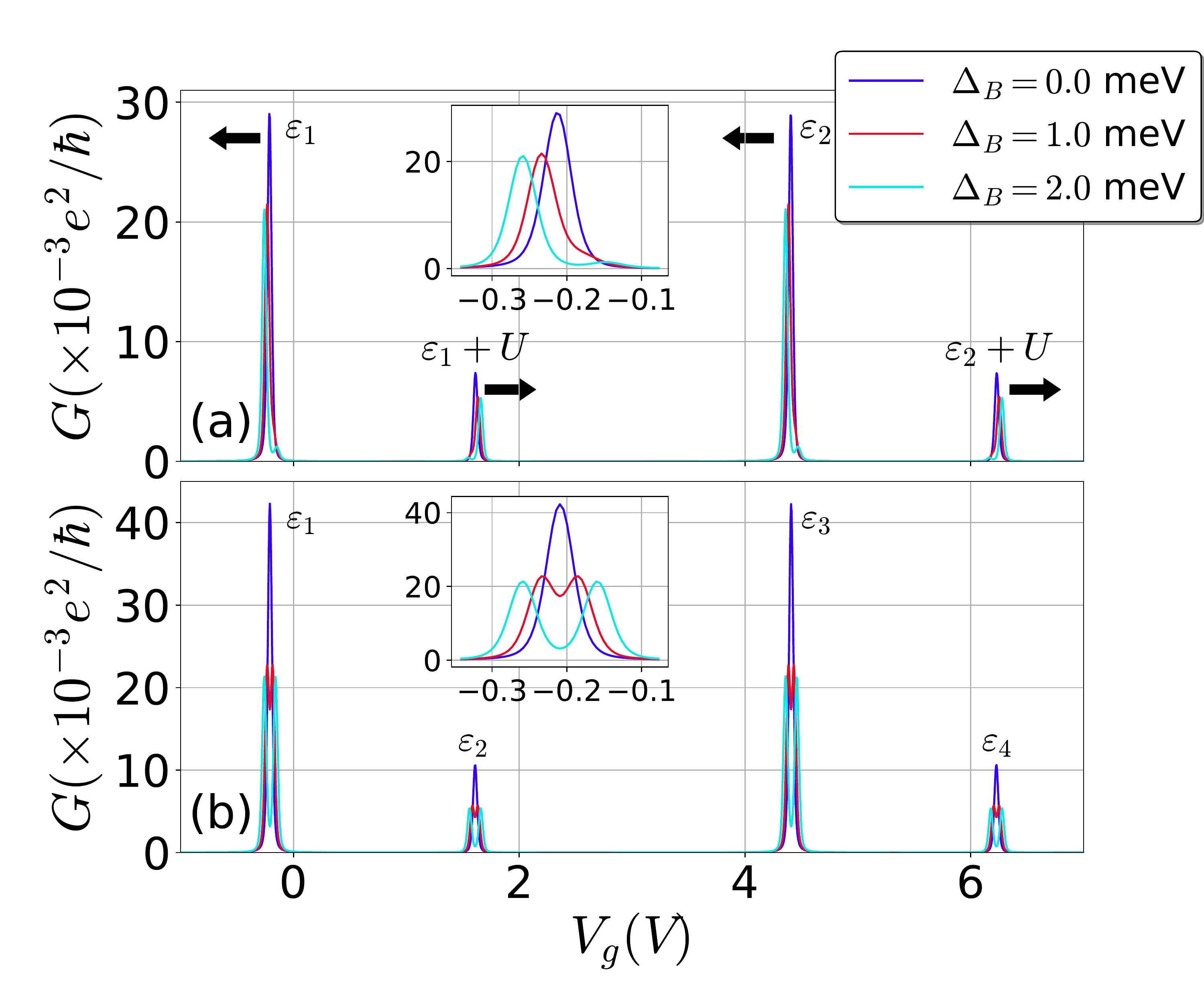}
  \caption{(a) Conductance $G$ at $T= 4$~K as a function of the gate voltage $V_g$ for different values of the Zeeman splitting $\Delta_B$ in the interacting model. We take two molecular levels with spacing $\Delta=140$~meV ~\cite{alv115},
each split due to Coulomb repulsion.
The couplings for the second level $\varepsilon_2$ are taken the same
as for $\varepsilon_1$.
Black arrows show how the peaks shift with increasing $\Delta_B$. (b) Same as (a) but in the noninteracting model with four single-particle resonances, which spin split as $\Delta_B$ enhances.
The couplings for the third (fourth) level $\varepsilon_3$
($\varepsilon_4$) are taken the same
as for $\varepsilon_1$ ($\varepsilon_2$).
Insets: leftmost conductance peaks
from the main panels are zoomed in for better visualization.}
  \label{fig:4}
 \end{figure}

An external magnetic field $B$ induces a Zeeman interaction in the molecule.
In the interacting model, this implies that Eq.~\eqref{Eq:mud} becomes~\cite{mar01}
\begin{equation}
\label{Eq:mudB}\mu_m(N) = \frac{(2N-1)e^2}{2C}-e \frac{C_LV_L+C_RV_L+C_gV_g}{C}+\varepsilon_N+\Delta_S g\mu_B B \, ,
\end{equation}
where $\Delta_S=S_z^N-S_z^{N-1}$, with $S^N_z$ the total spin of the molecule. 
Clearly, if the spin raises when adding an electron ($\Delta_S>0$), this leads to a negative
voltage shift of the peak associated to $\mu_m(N)$ (because $V_g$ must decrease to compensate),
while if the spin lowers ($\Delta_S<0$) the peak shifts to higher voltage values
with increasing $B$ (we assume a positive $g$ factor; for negative $g$ as in GaAs dots, see Ref.~\cite{duncan}).
Our argument then shows that in the interacting case
two consecutive conductance peaks will alternatively attract or repel each other.
For a given energy level the peak separation increases
as $U+2\Delta_B$, where $2\Delta_B=g\mu_BB $ is the Zeeman splitting,
whereas for different energy levels the peak separation decreases as $\Delta+U-2\Delta_B$.
In stark contrast, in the noninteracting model
each peak corresponds to a given energy level
[four levels displayed in Fig.~\ref{fig:4}(b)],
which split symmetrically under the action of a magnetic field (Zeeman splitting).
Therefore, a magnetic field makes it possible to
distinguish between two molecular resonances separated by either charging effects or
quantum confinement.

The differing response to the presence of magnetic fields leads to distinct conductance curves
for the interacting and the noninteracting cases.
In the interacting theory, the level $\varepsilon_m$ in Eq.~\eqref{Eq:Gr} becomes spin dependent,
$\varepsilon_{m\sigma}=\varepsilon_m+s\Delta_B$ with $s=+$ ($-$) for $\sigma=\uparrow$ ($\downarrow$). In the noninteracting
approach, both molecular levels in Eq.~\eqref{Eq:Tlandauer} become spin dependent as
$\varepsilon_i \rightarrow \varepsilon_{i \sigma}$.
The peak amplitude depends on $T$ and the particular choice of $\gamma$ and is thus sample-dependent (e.g., in the Electronic Supplementary Information$^\dag$ we show
that a different choice of $\gamma$ causes
the lower resonance to have a smaller amplitude compared to the Coulomb-shifted one, in contrast to Fig.~\ref{fig:4}).

The effects of the magnetic field in both models are shown in Fig.~\ref{fig:4}. We depict the electric conductance $G=(dI/dV)|_{V=0}$ as a function of the gate voltage for different values of the Zeeman strength $\Delta_B$.
For interacting particles [Fig.~\ref{fig:4}(a) for two energy levels $\varepsilon_1$ and $\varepsilon_2$ with interaction] the $G$ peak separation expands or shrinks as $\Delta_B$ increases, as anticipated above. This is a clear manifestation of a Coulomb-blockade split resonance. Contrary to this case,
the interacting model [Fig.~\ref{fig:4}(b) for four energy levels $\varepsilon_1$, $\varepsilon_2$, $\varepsilon_3$ and $\varepsilon_4$
without interaction] leads to
peak splittings when $B$ is such that $\Delta_B\simeq \Gamma$.

\begin{figure}[t]
\centering
  \includegraphics[scale=0.4,clip]{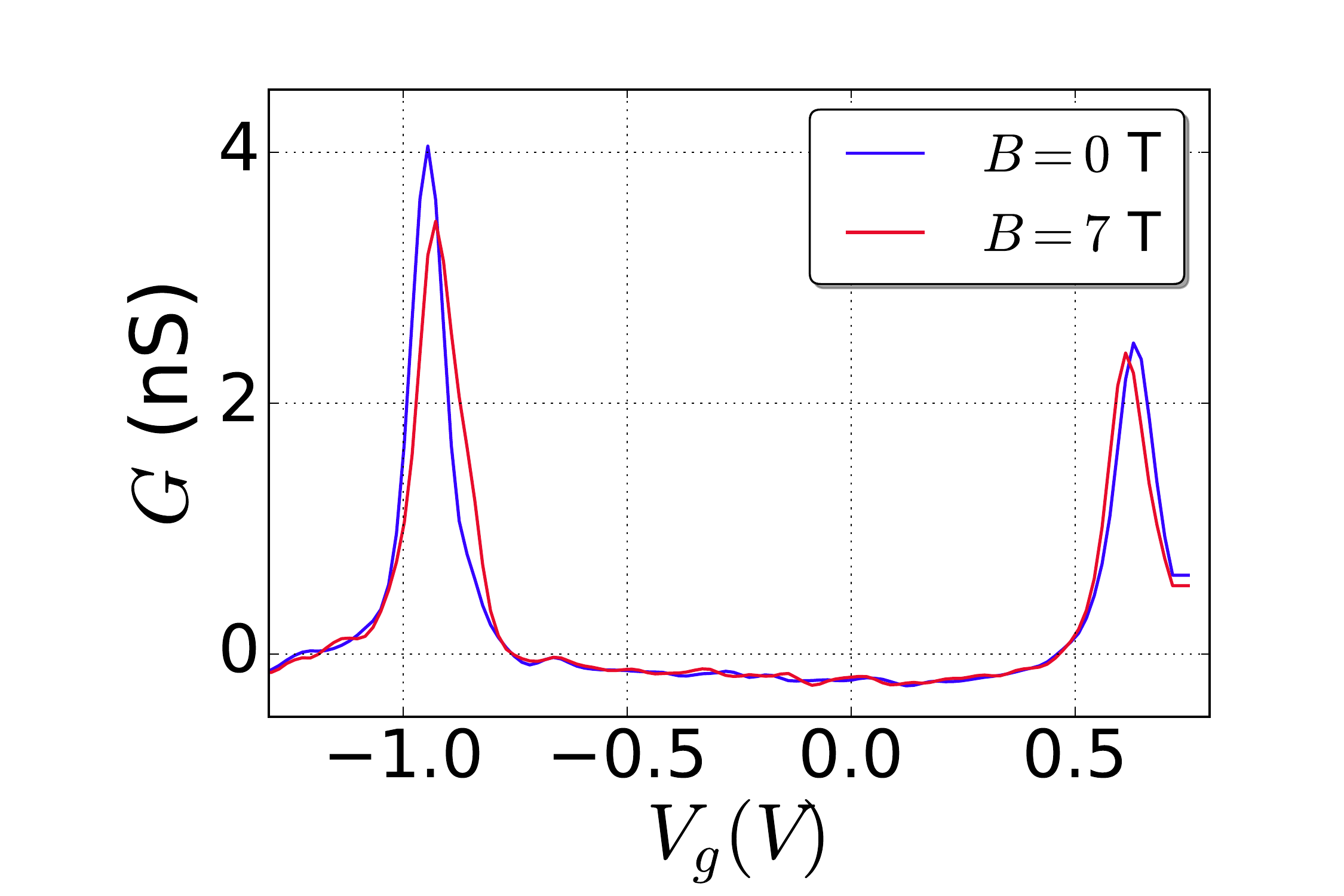}
  \caption{Experimental differential conductance measurement of a ferrocene-based single molecular tunnel junction for two values of the magnetic field $B$ at a voltage bias of $|V_{sd}|=80$ mV. Clearly, both conductance peaks
  are shifted as $B$ increases, which suggests that charging effects are the main transport mechanism.}
  \label{fig_exp}
 \end{figure}
 
To test our theoretical predictions, a room temperature electromigration-breaking three-terminal single-electron transistor with Au bias-drain electrodes and Al/Al$_2$O$_3$ back gates was used to study an individual S-(CH$_2$)$_4$-ferrocenyl-(CH$_2$)$_4$-S molecule [Fig.~\ref{fig:sketch}(c)] at low temperature (see Ref.\ \cite{alv115} for details on device fabrication).
Note that the molecules investigated in Ref.~\cite{alv115} were not tested with a magnetic field since its effect would not be observable at the lowest temperature (80~K) employed in that study.
Figure~\ref{fig_exp} shows the differential conductance as a function of gate voltage through the ferrocene-based single molecule junction obtained in the absence of magnetic field (blue data) and with a 7-Tesla field applied (red data). The measurements were obtained at $T= 4$~K with a bias voltage of 80~mV. Two peaks are clearly visible at $V_g = -1.95$~V and $0.65$~V. Upon application of a magnetic field, the peaks shift in voltage approaching each other
(see the stability plot in the Electronic Supplementary Information$^\dag$).
One possible scenario is that the magnetic field drags the ferrocene unit slightly, hence distorting the molecule and changing its coupling to the electrodes and gate. However, it is unlikely that this effect leads to such a large change in potential energy
since it would change the transport
excitation slopes, which is not observed in our data.
Additionally, strain-induced changes in the electrostatic coupling of the molecule will not discriminate levels according to its spin value. We next argue that the most natural explanation is in terms of charging effects.

We first notice that the peaks in Fig.~\ref{fig_exp} do not Zeeman split under the action of the magnetic field, as it would be expected from two distinct charge degeneracy points arising from two different molecular levels. Instead, the peaks shift, conferring them a Coulomb blockade nature resulting from strong charging energy effects (as explained above). The gate voltage shifts are actually pretty similar for both peaks. For the left peak in Fig.~\ref{fig_exp} is $+ 18$~mV, while for the right peak is $-19$~mV. Indeed, using the coupling capacitances of the molecule with the respective electrodes, extracted from measurements at different bias voltages, the observed shift in gate voltage for each peak gives $\Delta_S = -1/2$ for the left peak in Fig.~\ref{fig_exp} and $\Delta_S = +1/2$ for the right peak (i.e. a net spin change of $\frac{1}{2}$ as it corresponds to adding an electron into the molecule). However, the peaks do not repel from each other in increasing magnetic field, as would be expected if originated from the same Coulomb-blockade energy level splitting. This means that the two observed peaks belong to two different Coulomb-blockaded energy levels, whose corresponding pairs lie beyond the measuring gate voltage window in this measurement. It is very likely that
the associated charge states from left to right in Fig.~\ref{fig_exp} correspond to
Fe$^{3+}$, Fe$^{2+}$ and Fe$^{+1}$. (Note that the Fe$^{+1}$ state can be observed at low temperatures similarly to the work of de Leon \textit{et al.} in Ref.~\cite{lea08}.)
In any case, the observation of the gate voltage shift under the action of a magnetic field allows the unequivocal association of the differential conductance peaks to charging effects, and showcases a magnetic field as a powerful tool to determine the nature of transport excitations in single-molecule junctions.  

\section{Conclusions} 
In conclusion, we have shown that the temperature-dependent transport properties of molecular tunnel junctions
can be explained using a fully interacting model that takes into account Coulomb blockade effects.
Since a noninteracting theory with two molecular levels also agrees with the experimental data,
it is natural to ask what the nature of a real resonance is. We have suggested that
an externally applied magnetic field can be used as a transport spectroscopy tool that distinguishes
between the two models. Employing values extracted from the experiment,
we find that the conductance peaks of a molecular bridge
with Coulomb interactions
shift as the magnetic field increases while in the noninteracting case the transmission resonances are all Zeeman split.
Finally, we report on conductance measurements in the presence of a magnetic field
that point to Coulomb-blockaded resonances in a ferrocene-based molecular tunnel junction.
Since charging energies are expected to be similar in molecular tunnel junctions,
the interaction model is likely to apply in most transport experiments where molecules are in the weak coupling regime.
In our case, the ferrocenyl units are weakly coupled to the electrodes due to the insulating character of the alkyl chains.
Our work thus represents an important contribution that will help identify the transport regime of molecular junction experiments.

\section*{Acknowledgments}
We acknowledge support from MINECO under grant No.~FIS2014-52564, the CAIB PhD program, the National Science Foundation under grants NSF-ECCS Nos.~1402990 and 1518863 and the Ministry of Education (MOE) under award No.~MOE2015-T2-1-050. Prime Minister's Office, Singapore, under its Medium sized center program is also acknowledged for supporting this research.





\bibliography{rsc} 
\bibliographystyle{rsc} 

\end{document}



\title{Electronic supplementary information: How to distinguish between interacting and noninteracting molecules in tunnel junctions}

\author{Miguel A. Sierra}
\affiliation{Institute for Cross-Disciplinary Physics and Complex Systems IFISC (UIB-CSIC), Palma de Mallorca, Spain}
\author{David S{\'a}nchez}
\affiliation{Institute for Cross-Disciplinary Physics and Complex Systems IFISC (UIB-CSIC), Palma de Mallorca, Spain}
\author{Alvar R. Garrigues}
\affiliation{Department of Physics, University of Central Florida, Orlando, Florida, USA}
\author{Enrique del Barco}
\affiliation{Department of Physics, University of Central Florida, Orlando, Florida, USA}
\author{Lejia Wang}
\affiliation{School of Chemical Engineering, Ningbo University of Technology, Ningbo, Zhejiang, 315016, P.R. China}
\affiliation{Department of Chemistry, National University of Singapore, Singapore}
\author{Christian A. Nijhuis}
\affiliation{Department of Chemistry, National University of Singapore, Singapore}
\affiliation{Centre for Advanced 2D Materials, National University of Singapore, Singapore}

\maketitle

\section{Additional measurements} 

The experiments were done for a fixed bias near zero and by sweeping the gate voltage continuously in order to increase the definition of the peaks associated to crossing the charge degeneracy points. This is standard procedure when checking if there are molecules present in the nano-transistors, and we used it here to minimize the time between measurements with and without magnetic field. Molecules in electromigrated three-terminal junctions are very unstable, and frequently move, changing their coupling to the transistor leads. This was actually the case of the molecule measured here.

We have measurements of the diamond for one of the charge degeneracy points of this molecule (see Fig.~1), where one can see how all excitations are equally affected by a magnetic field. Although not with the same precision than in the measurements for a single bias potential, the shift can be clearly resolved in these results, which show that the shift affects equally all excitations and that there is no Zeeman splitting (at least not comparable to the observed shift). Note that for these measurements, the molecule has already moved with respect to the measurements presented in the main text, and the first charge degeneracy point appears now at around $V_g=-1.5$~V for both fields ($-0.95$~V in the measurements included in the main text).

\begin{figure}[h]
\centering
  \includegraphics[width=0.49\textwidth,clip]{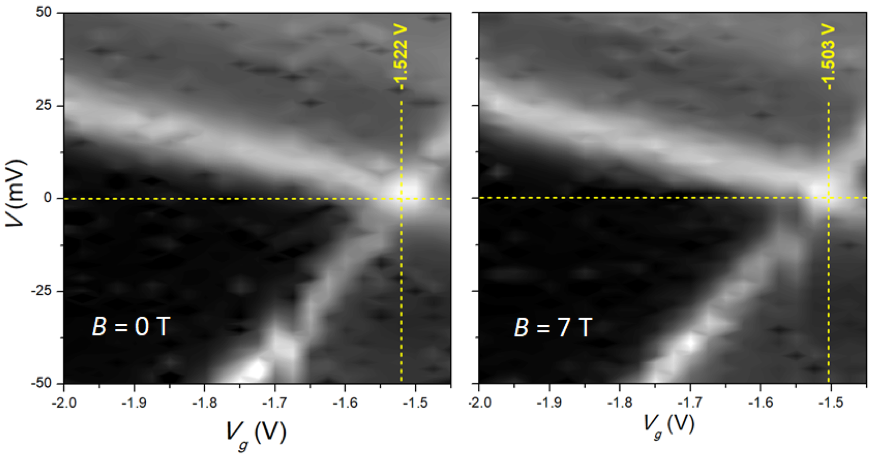}
\caption{Excitations in the presence of a magnetic field.}
\end{figure}

\section{Additional calculations}

The detailed amplitude ratio depends on the molecule’s coupling to the metallic reservoirs. For $\Gamma\ll k_BT$ the peak height is proportional to $\Gamma_1\Gamma_2$ and inversely proportional to temperature~\cite{been}. In general, the couplings are Porter-Thomas distributed and therefore the peak amplitudes fluctuate. Below, we present in Fig.~2 theoretical conductance curves where we used different values of $\gamma$. Our results show that the lower resonance has a smaller amplitude compared to the Coulomb-shifted one. 

\begin{figure}[h]
\centering
  \includegraphics[width=0.49\textwidth,clip]{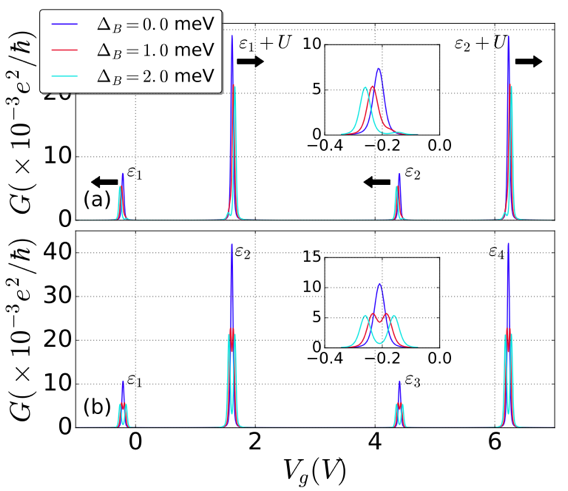}
\caption{Conductance curves for different values of the tunnel couplings: 
$\gamma_{L1} = 0.01$~meV, $\gamma_{R1} = 0.4$~meV,
$\gamma_{L2} = 0.05$~meV and $\gamma_{R2} = 0.4$~meV.
We use the couplings for $\varepsilon_2$ in (a) the same as for $\varepsilon_1$ while in (b) the couplings for the third (fourth) level $\varepsilon_3$
($\varepsilon_4$) are taken the same
as for $\varepsilon_1$ ($\varepsilon_2$).}
\end{figure}